\documentstyle[prl,aps,twocolumn]{revtex}
\def\break#1{\pagebreak \vspace*{#1}}
\begin{document}

\draft
\pagestyle{empty}

\title{ ESSAY ON MESOSCOPIC AND QUANTUM BRAIN}
\author{Haret C. Rosu 
}
\address{ \begin{center}
Instituto de F\'{\i}sica de la Universidad de Guanajuato, Apdo Postal
E-143, Le\'on, Gto, M\'exico\\
Institute of Gravitation and Space Sciences, Bucharest, Romania\\
rosu@ifug3.ugto.mx\\ 
(received: February 7, 1997; published on March 1, 1997 by
 Metaphysical Review 3 (8), pp 1-12)
\end{center}
}

\maketitle
\widetext
{\scriptsize The fox knows many things,
\hspace{4.05cm}
Brooks has used the subsumption architecture to build insect-like robots.} 

{\scriptsize But the hedgehog knows one big thing,\hspace{2.8cm}
But insect minds are not very interesting.}

{\scriptsize Archilochus \hspace{6.1cm}
We are now exploring the space between the insect and the adult human.}

{\scriptsize ... what about the grasshopper ? \hspace{3.45cm}
D.C. Dennett, Phil. Trans. R. Soc. Lond. A {\bf 349}, 146 (1994)}

\vskip 0.5cm

\begin{abstract}
In the pure essay style (no mathematical formulas), I present a number of
speculative reflections and suggestions on possible applications of
mesoscopic methods and of
quantum mechanical concepts to as such a complex system as the human brain.
As an initial guide for this essay I used ``The Emperor's New Mind" of
Roger Penrose.

\end{abstract}


\narrowtext
























\section{Introduction}

The almost one hundred years of historical development of quantum theory are
a manifest
proof of its viability and successfulness, despite a number of persisting
conceptual and/or philosophical difficulties, e.g., measurement, quantum-Zeno,
and EPR paradoxes, that may be considered ever-lasting open problems.
Due to their versatility
the quantum methods can be applied in principle to any
space-time scale, when amended with corresponding innovations, usually by
generalizing certain delicate interpretational aspects. For example,
one may encounter ambitious programs such as describing the whole
universe in quantum mechanical terms, a case in which the usual
Copenhagen interpretation, apparently sufficient at microscales,
have to be replaced by more general schemes, as for example,
the ``sum over histories" interpretation \cite{gmh90},
a modern variant of Everett's ``relative state" (1957) \cite{ev}, or of the
slightly different language of ``many worlds" \cite{bsd70}. For a recent `map'
of the various interpretations and other issues of quantum mechanics, I
recommend the paper of Sonego \cite{s92}.  Unfortunately,
what happens when one is trying to extend too much the usual domain of a
theory, even if it is of the rank of quantum mechanics, which is
{\em superb and useful} in Penrose's classification, is
to turn it into a purely formal and almost unuseful scheme.

Since a common way of scientific reasoning in physics is that
phenomena at normal macroscopic scales are to be explained in terms of
concepts
built up of quantities formally existing at microscopic scales, many people
believe that quantum theory is a universal theory \cite{pz82}.
Therefore quantum
theory/mechanics should have something to say regarding one of the
most sophisticated systems, and actually for the time being,
\break{1.9in}
the most
sophisticated we know about, which undeniably
is the {\em human brain}.
This ``porridge-like" biological assemble is the command unit of the
human body and of extreme importance to all of us for any need, including the
scientific one.
One can think of it at three spatial scales: the microscopic, the mesoscopic,
and the macroscopic ones. By microscopic scales I would like to mean quantum
length scales,
i.e., $10^{-10}-10^{-9}$ m, the mesoscopic scales, where according to
Feynman \cite{fe59} ``there's plenty of room...", are those between
$10^{-9}-10^{-7}$ m, while beyond that one can say that we passed into the
macroscopic realm, which in fact for the brain reduces to the centimeter scale.
This division is of course not sharp and there are no
well-established criteria for the relative separation. Most of the brain
activity proceeds at the mesoscopic and macroscopic scales and it seems
a priori unuseful to think of quantum
features and quantum mechanics for such a complex self-organization.
But for a physicist this is not so, and
as a matter of fact, he/she should attempt at finding arguments for making
relevant the quantum features of the human brain. Moreover, 
some of the quantum methods and ideas can find interesting applications in
this field even at scales which are not properly quantum ones. The spatial
scales mentioned above are standard ones in physics, i.e., they are the scales
with which most of the physicists are dealing.
Since human brain is a complex physical, biological, and information-processing
system, one will expect multiple spatial and temporal scales to be mixed up,
with interactions taking place at multiple hierarchical levels. Therefore
the structural division of the brain activity usually
considered by neuroscientists might look more natural, i.e., the microscopic
scales are those
of synaptic-neuronal interactions, the mesoscopic ones belong to minicolumns
and
macrocolumns of neurons, and the macroscopic scales are characterized by the
regional activity over centimeters of neocortex (see \cite{ing} and the next
section). The columns are defined as filamentary cluster structures of neurons 
in the (neo)cortex.

In the following, the reader will find several incipient and quite provisional
opinions on the problem of the human brain at the mesoscopic and quantum
level
that I started to gather together mainly in the summer of 1992, when I began
this essay while spending some really good time browsing
in the ICTP-SISSA libraries and looking more carefully into {\em The Emperor's
New Mind}. Being an essay, I
escape any mathematical rigor, thus allowing me to utter, even though in a
cursory manner, what might be some
interesting and hopefully useful ideas for future analyses.

Previously to start reading this essay, I recommend the reader to take a
look in Chapters 9 and 10, at least, in the aforementioned book of 
Penrose \cite{pe89} for world-wide known opinions on the brain, to which I will
frequently refer in the following.
Philosopher Owen Flanagan has recently classified many of the scientists not
belonging
to the mainstream neuroscience as ``the new mysterians". These are supposed
to be people
whose more or less declared beliefs are that topics such as
{\em consciousness} and {\em free will} are too profound for scientific
 studies. Therefore what ``the new mysterians" do is
to mistify (consciously or unconsciously) those concepts, usually by
relating them to other
mysteries (of quantum mechanics in the case of Roger Penrose; recall that
Chapter 6 in {\em Emperor's} has the title: {\em Quantum magic and quantum
mistery}). I am negative
to such an opinion, and I found in expressing my disagreement another
motivation for the present essay.
I think that scientists have the right to speculate.
It is only a question of time for some small amount of their speculations to
convert into scientific and even technologic truths. These were the main
underlying arguments for writing the present essay.

\section{What is the human brain ?}

The brain is by itself a complex, i.e., self-organized
quantum-meso-macro-scopic system/state of biological material which is more
than a logical machine (composite- computer), showing some ability to react
at phase correlations. J.J. Hopfield \cite{lh82} remarked that the question
``How does it
work ?" is one of the best motivations for many scientists. In the case of
the brain, an efficient answer is ``it is doing computations", and this in
its particular ``biological" way.  The powerful paradigm here is to view the
brain composite-computers as input -output devices performing transformations
on the input signals to generate the output ones. However, this in-out mapping
is extremely complicated in the case of biological computers. It is also the
main subject for the artificial intelligence projects. Apparently, the paradigm
of computing is at odd only with the concept of consciousness of the human
brain. If a measure or a parametrization of the consciousness will be found,
for human brains as well as for all biological computers, then this will make
the difference between biological computers and electronic ones.

One defines a central nervous system to be a network
of $\cal N$ interconnected neurons. The total number of neurons is
approximately $10^{10}$, each of which connects to a so called signal target
 $\cal S{\cal T}$ made of a cluster of some
$10^{3}-10^{4}$ neurons.
 The nearest neighbour connections are called
synapses ({\em Sy}), by which neurons are sending electrical and
chemical signals to their $\cal S{\cal T}$ cluster. It is supposed that
any {\em Sy}
is in one of the two possible states: firing and non-firing.
As such, a certain analogy with spin systems, the well-known Little-Hopfield
model \cite{lh82},
has been developed for simulating the associative memory of neural networks
\cite{tk} and constructing learning algorithms for artificial intelligence.
I would like to suggest possible connections of the activity of neural 
networks with some of the self-organized criticality models, that clearly one
can envision,
especially with the forest-fire (FF) class of models \cite{bct90}.
By appropriately
generalizing the automaton rules of the FF models one can put them in
correspondence with the quasi-stable patterns (memory) of neuronal firing
activity.
On the other hand, the Hopfield model is based on a quasispin representation of
the physical states of
firing and nonfiring \cite{MCP43}: memories to be stored are just patterns of
binary sequences of quasispin variables $S_{i}=\pm 1$ where the index $i$
is running over the whole number $\cal N$ of neurons in the network.
Such a sequence may be regarded as an $\cal N$-component vector,
characterizing the patterns, which are stored if they are turned into
attractors of the spin-flip dynamics. This dynamics is governed by the signs
of the exchange sums, where the coupling constants are considered to be the
synaptic strengths. One can turn given patterns into attractors by the Hebb's
mechanism \cite{h49}, i.e., by appropriate modifications of the synaptic
strengths, known as learning algorithms. Major difficulties were surpassed by
bounding the synaptic strengths (learning within bounds), and at the present
time the ``Ising-like" models with all the apparatus of spin-glass theory
\cite{sg} are by far the most powerful paradigm of
physics for studying the brain activity considered as sets of computations.
These are, in a few words only, the basic facts required in order to proceed
in a constructive- computing manner towards further understanding of the
higher functions of human brain and/or the interconnections among its
subsystems (visual cortex, somatosensory cortex, motor cortex, thalamus,
peripheral cortex).

Hopfield's paradigm is fine and quite efficient for artificial intelligence
purposes.
Nevertheless there is one really difficult question for it and this is the
title of the first chapter in {\em Emperor's}:
{\em Can a computer have a mind ?} In other words, what is the fact
providing
the distinction between a biological computer and an electronic one ?
Is a biological computer just a more complicated electronic one or is
there a fundamental difference ? Is this difference provided by quantum
mechanics ?
I shall try to formulate some arguments based on quantum
ideas in Section IV below. Here I shall list other general properties of the
brain that one can notice when is passing at the level of the higher brain
functions:

(i) The higher functions are in general delocalized, display some degree of
stochasticity, and are intercorrelated in parallel computing manner. There are
many unresolved questions concerning the integration of cortical activity and
the `higher' integrative areas \cite{cr84}

(ii) The neurons have the capacity of working out several inputs and are
selecting the output signal and its frequency, the cooperative result of
such a processing being a kind of generalized holographic recording of the
outside world.

An interesting columnar self-organization of the neocortex is well-known:
``minicolumns " of about 110 neurons (about 220 in the visual cortex) comprise
modular units vertically oriented relative to the warped and convoluted
neocortical surface through almost all the regions of the neocortex.
The short-ranged fiber interactions (both excitatory and inhibitory) between
neurons take place within about 1 mm,
which is the extent of a ``macrocolumn" comprising about one thousand
minicolumns, whereas the long-ranged cortico-cortical excitatory fibers
(the white matter) have an averaged length of several centimeters.
This structural organization supports the idea of
computing-oriented activity of the brain.

Shelepin, whom I cite in Section IV below, has suggested the theory of
complex Markov chains as a
sufficiently general mathematical description of the higher functions of the
brain, which include quantum mechanics as a particular case, but in any case,
one has to be aware of the impressive panoply of disciplines contributing to
their understanding: neurobiology, computer science, biochemistry, artificial
intelligence, molecular biology, mathematics, psychology, physics, and
philosophy.

\section{Consciousness and mesoscopia}

Perhaps the most fundamental notion in neuropshycology is the
global attribute of the brain known as consciousness.
In general terms, what we usually call awareness or consciousness or
``unique personality" might be considered a problem of spatio-temporal
synchronization between
the two cerebral hemispheres. This interpretation comes out from an
interesting neuro-disease, which manifests itself
by the so-called ``multiple personalities'' cases to be found  for example
in the book
of Gazzaniga and LeDoux \cite{gl11}. This neuro-disease is the result
of the therapeutic operations (severing of the corpus callosum) for some
forms of epilepsy, and more generally can be considered as split-brain
experiments. Such cases have the exterior data mapped
only on one cerebral hemisphere without the other hemisphere being aware of
them. Thus, one can think simply of a desynchronization of the
two hemispheres at the level of their neuronal signals. This alone explains
the attention paid to the synchronized oscillations in the cerebral cortex
\cite{ll88}. {\em Emperors's} p. 385 mentions also the interesting `P.S.'
split-brain case revealed by neurophysiologists, showing a transient phase
in which only one hemisphere could speak, but both hemispheres could
comprehend speech. For the cases with removed portions of visual cortex
and comments on the phenomenon of {\em blindsight} as related to consciousness,
see {\em Emperor's} pp. 386-387.

Clearly, it is extremely difficult to accept a definite physical base for such
an esoteric concept as consciousness dealing mainly with the subjective
activity of the brain. It may be called a sense for which the receptive organs
are directly the neurons, in which all the other sense stimuli can be more or
less included on a subjective base, that is with degrees of importance varying
from one brain to another. The neuronal global response to such a brain
activity is the personal
representation of the exterior and interior world altogether and may be called
consciousness. It is also a parameter of the evolution in time of an
individual brain, obviously connected with both short-term and long-term memory.
It is a direct neuronal ``pshycological", and sometimes almost physiological
sense that occurs as an outcome of all the mental functions
of an individual brain working in {\em synergis}, and probably, from this
standpoint, one can interpret it as an informational measure of the coupling
between the `subjectivity' and the `objectivity' of a brain.

There are at least two physical phenomena contributing to consciousness in its
objective form. One
is the synchrony of the neurons. When synchrony is between the neurons
of the two hemispheres it provides the `unique personality' character
of the brain. The other mechanism is the stationarity of the 40 Hz collective
oscillations of the neurons as shown by experiments on animals. Synchrony and
the 40 Hz oscillations together are related to the so-called `binding problem'
in neuroscience which is essentially the making of a unified perception.
But what makes neurons to oscillate collectivelly at roughly 40 Hz. Is this
a reflection of the nonlinear dynamics of the neuronal network as a result of
functions such as memory and attention or it has to do with the microtubule
architecture of the neuron skeletons ? Again cummulative effects can be
invoked. The microtubules, which are long (350-750 microns in the axons), 
and rigid polymers made of a globular protein called tubulin, were suggested to
generate quantum effects of importance for consciousness by Penrose \cite{p94}.
I would like to come here with an argument of interest for microtubules
taken
from the mesoscopic phenomena recently put into evidence in the realm of
carbon nanotubes (for their history see \cite{g92}). Carbon nanotubes are
thread-like structures forming in carbon deposition stimulated by an electron
beam, and are pretty well observed in scanning transmission electron
microscopy \cite{n92}, and, as a matter of fact, they are amongst the few
laboratory-produced structures covering
the crossover from microscopic to the mesoscopic regime. In an interesting
experiment, Kasumov, Kislov and
Khodos \cite{kkk92} observed displacements of the free ends of threads of
amorphous hydrocarbons of 200-500 $\AA$  in width and 0.2-2.0 $\mu$m in length
relative to a fixed reference point on the screen of a transmission electron
microscope. The minimal displacements were of about 5 $\AA$, and the
observations
were made in a stationary regime of the threads, i.e., very low density of
the beam current (0.1 pA/cm$^{2}$). They observed random jumps of the free ends
of the carbon threads of 10-30 $\AA$  in length with a frequency of 1 Hz. All
the possible reasons of induced vibrations were taken into account by the
authors with the conclusion that
no classical external force can explain the jumps and finally they attributed
the oscillations to jumping effects related to spontaneous localization
ideas of Ghirardi, Rimini, and Weber \cite{grw86}. In our opinion, the jumps
in length of
the carbon nanotubes can result from a mesoscopic Brownian
motion in which there is a competition between some dynamical instability
and damping, being different from the microscopic Brownian
jumps which never damp out. If such jumps will be confirmed by other
experiments, and their origin identified, there will be important consequences
for neuronal microtubules too. For instance, one can associate
the 40 Hz oscillations either with the frequency of the jumps of
the network of neuronal microtubules due to a mesoscopic Brownian motion as
mentioned above or with spontaneous localization
ideas \cite{grw86} as applied to microtubules.
Actually, microtubules are already an active experimental
and theoretical research field \cite{d84}. Their interesting growth properties
have been recently under focus \cite{dl93}, and also non-linear energy-transfer
mechanisms in microtubules have been proposed \cite{stz93} making the field
more physical. They may play an important role in the brain plasticity
({\em Emperor's} pp. 396-398). At the same time, it is quite obvious that
graphene
tubules can reveal many phenomena of worth for biological microtubules as well.

\section{Hints for quantum approaches to the human brain}

I shall start this section by recalling Penrose's rather strong speculation
on the existence of single-quantum sensitive neurons ({\em Emperor's}
pp 400-401). Yet independently of this speculation, there are various other
ideas concerning possible quantum treatments of the brain.

I would like to present shortly some facts from superfluorescence (SF)
that might be of importance for Hopfield's paradigm as I already mentioned
at the end of Section II.
Perhaps the simplest and probably useful way to think of
quantum effects within human brain is to consider it as a kind of generalized
Dicke superfluorescent (superradiant) system. This has been suggested by
Shelepin \cite{she88}
as an analogy for the two-position switch of axons. Four decades ago, Dicke has
pointed out that $\cal N$ atomic oscillators interacting with a common
radiation field are not independent and live in a correlated state
\cite{di54} that, under certain conditions, can display a collective radiative
deexcitation, with all $\cal N$ oscillators acting like a single rigid dipole.
In the original treatment, the matter-radiation system is
described by a Hamiltonian of three terms corresponding to a collection of
two-level atoms, a one-mode field, and a one-photon Dicke interaction (a
simple coupling between the transition operators and the absorption/emission
operators of the photon).
On these lines, particularly interesting would be to reveal
counterpropagating correlations of the type recently put into evidence and
discussed in solid-state superfluorescence \cite{j} \cite{b} with
quasi-one-dimensional active volumes (pencil-shaped excitation volume) of length
much longer than the emitted wavelength. Let me point out that even of more 
relevance to the problem of superradiant neurons
is the observation of {\em hyperradiance} (HR) from phase-locked soliton 
oscillators
in the setup of {\em long} Josephson junctions \cite{hyp}, because neurons are
closer to soliton oscillators than to atomic ones. In any case, the
{\em hyperradiance} phenomenon must be investigated in detail
in the newly fabricated superconducting neural circuits \cite{m93}.
To pass to neurons, one can simply assume
that SF brain phenomena are induced by certain particular neurons
acting similarly to the SF centres in crystals, whereas one can invoke some
magnetic coupling between the synapses when the analogy with the Josephson
junctions is pursued. In the first case for example, one is
allowed to consider distributed-feedback structures due to density fluctuations
of the SF neurons as the origin of the correlations.

Perhaps it is worthwile to note that the strong correlations
between counterpropagating one-dimensional pulses are absent in the gas phase.
One might have in this way more than a na\"{\i}ve answer to the na\"{\i}ve
question
of why the brain is in a solid-state phase and not in a gas one. Clearly, it
would be extremely interesting to look for counterpropagating
correlations between the two cerebral hemispheres and to see the
implications for brain synchronization. Their similarity with the EPR
quantum correlations \cite{h11}
should be investigated in order to get insight and provide good answers
to the question: {\em Does quantum mechanics/quantum-like effects make us
intelligent ?} It is worth mentioning at this point that some time
ago, Vindu\v{s}ka \cite{v91} elaborated on the impossibility of creating
quantum correlations with electronic computers. It might well be that a
biological computer makes use of EPR-type correlations, thus promoting itself
to a superior level of existence.
What one should keep clear in his mind is that superfluorescence is a
cooperative phenomenon, i.e., the output is proportional to the squared
number of neurons involved, and it is due to some type of emission process
and not to an amplification of an input signal. This implies a ``laser"-like
action of some brain activity.

On the other hand, there are many mathematical aspects involved in treating
the human brain
as a macroscopic quantum state. The first problem is to define
rigorously the macroscopic brain quantum state. In this respect, we draw 
attention to the paper of Duffield, Roos, and
Werner \cite{drw}, who defined some notions of mean field limit
for nets of states converging to a macroscopic limit state.

Of much relevance to the field of neuropsychology might be the 
experimental findings of Kelso {\em et al} \cite{k11} who put into evidence,
by means of SQUID detectors, spontaneous transitions in the
neuromagnetic field patterns. They claimed that such transitions are to
be associated with the switching of the non-equilibrium patterns
formed by the brain during the transition between
coherent states, and so from one behavior to another one. One might
guess that various types of coherent and squeezed states \cite{t94}, when
appropriately generalized, and
within information-theoretic pictures \cite{ho94}, will have important
applications in this field.

\section{Quantum effects in human receptors}

We are interested in the human receptory organs since they are the places
where manifestations of quantum effects from the standpoint of their
sensitivity and response have been reported so far.
At the cell scale, human brain has quantum (molecular)
receptors of the outside fields. These receptors absorb
electromagnetic radiation
at the level of tens to thousands of quanta per mode as well as phonons in
the same amount. More powerful fluxes are already damaging.

\subsection{Visual or electromagnetic reception}

Perhaps, the best sensory system in which one may have hopes for studying
quantum correlation phenomena
to be associated with the human brain is the visual system (from the eye up
to the visual cortex). In fact, in this case one encounters experimental
results on the rod sensitivity to single photons. Actually,
biological photoreception has mesoscopic scale, and as such, is just
at the transition point from quantum reception to classical one.
For a good introduction to quantum fluctuations in the human vision we
refer to the review paper of M.A. Bouman {\em et al} \cite{b11}.
For the absorption of a single photon by a rhodopsin pigment and
its amplification ending up into a neural response see Lewis and
Del Priore \cite{lp11}, and for the responses of the retinal rods of toads
to single photons see Baylor, Lamb, and Yau \cite{bly79}. Penrose is also
citing Hecht, Shlaer, and Pirenne \cite{hsp41}, who established in a famous
experiment that an input
signal of seven photons is required by humans for conscient perception.

I now address the relationship between the electromagnetic vacuum
fluctuations and the possibility of four-dimensional and more-dimensional
vision. My point is that the electromagnetic zero-point fluctuations
are not very sensitive to the spatial dimensions of the macroscopic
world. In other words, the number of spatial dimensions is a quite free
parameter at the level of vacuum fluctuations \cite{nd}. Of course, the
conversion of two-dimensional images into three-dimensional ones is well
explained in the optics of the eye
as a stereoscopic effect and it is for this reason that we need two eyes,
but here I am referring to more-than-three spatial dimensions.
In my opinion, the Regge calculus approach \cite{re61}
to the more-dimensional manifolds, in its strict geometrical meaning,
will be quite useful for the problem of producing
vision in more dimensions, especially when the quantization of
4D Regge links will be properly understood \cite{Kh94}. 
The detailed features of the Regge quantum links will be essential in
proceeding toward a biological more-than-three dimensional vision. 
Moreover, one
should be aware of the experimental discovery of Hubel and Wiesel \cite{hw65}
who first observed that endstopped hypercomplex cells (that is, selective
to moving-bar stimuli of specific lengths) in the visual cortex
could respond to curved stimuli and sugessted they might be involved in the
detection of curvature.
More recently, Dobbins, Zucker and Cynader \cite{dzc87} provided both a
mathematical model relating endstopping to curvature and  physiological
evidence that endstopped cells in area 17 of the cat visual cortex are
selective for curvature.

There seems possible the implementation of multi- dimensional image
construction as well as multi- dimensional photoreceptors at the mesoscopic
level, either by using new types of ``depth" effects or holographic methods.
Also, more should be known on the connection between the internal
representations of rigid transformations and cortical activity paths as 
suggested by Carlton \cite{ca88}.

Let me remark on another important feature of living creatures. While within
the sonic world, the living creatures possess as a rule both receptory
and emitting organs, this is
not so in the electromagnetic world, where, in overwhelming majority, only
receptors are present, and there is no electromagnetic `mouth'. Moreover,
if this is
to exist, it should be a kind of biological laser \cite{f52}, in order to be
used for communication purposes. Although in the animal world there are
certain species of fishes possessing organs recepting and emitting electrical
pulses \cite{b94}, it appears that
the electric activity of the human brain, which is chemical in essence, is too
weak to sustain a lasing activity of the brain, at least of the electromagnetic
type. This looks frustrating, but we have to accept that it is much easier to
build up mechanical organs than laser ones using biological materials.

Finally, we recall that according to Chomski \cite{ch}, the
fisiology of the eye-brain system is essential in interpreting the
various trajectories we are observing in our visual field. Such an
argument is put forth as a consequence of the so-called ``rigidity
principle" in human vision, that is the interpretation of the visual
scene in terms of rigid objects in motion. On the other hand, the
animal visual systems are projected to react to other types of
movements.

\subsection{Hearing or sonic reception}

Quantum detection can be looked for in other sensory systems, in
particular in the hearing system, where by quantum one should mean the phonon,
although one can immediately estimate that the thermal environment actually
forbids single phonon detection for humans \cite{dw89}. In this
subsection I would like to draw attention to an ethnological claiming I
heard about in Trieste.
Some time ago, the ethnomusicologist Mantle Hood wrote an
essay on a ... quantum theory of music \cite{mh}. He advocates the
idea that a manifestation of Bohr complementarity principle is to
be encountered in this discipline of arts as {\em the continuity of the
first partial of a tone sounded and the discontinuity of constantly
shifting energies in the distribution of upper partials}.
These ethno-concepts are not clear to the present author who is merely
quoting the paper as a curiosity.
 According to Hood, {\em Musics}, as a form of cognitive learning,
is based on physiological responses to aural stimuli transcending
any mechanical differences
 in construction between the musical instruments. I remember that, during my
stay in Trieste, I participated in Prof. Hood's ethno-experiment,
which meant just hearing successively as diverse instruments as: Scotish
bagpipe, flute, tambura, mridangam, Tibet funeral horns, Korean kayagam,
Japanese gagaku, Irish tin whistle, and so on, in order to test his assumption,
but frankly I was not capable
of saying anything interesting about my aural stimuli.

As for the mesoscopic musical scales to which some technologies are already
knocking the door, one can foresee the numerous applications of wavelets in
processing musical sounds \cite{n94}. The wavelet approach \cite{chu} looks
already essential for studying the hearing system of the brain,
as well as the visual system and other brain phenomena, at the mesoscopic scale.

\subsection{Uncertainty principles}

The usage of wavelet principles is not at all new for psychophysical
experiments,
especially in models of vision. The old Gabor functions (harmonic
oscillations within Gaussian envelopes) \cite{g46} are in fact wavelets,
and have been introduced by applying arguments from quantum mechanics. Gabor
demonstrated that this class of ``modulated probability pulses" is optimal
in the sense that it possesses the smallest product of effective duration
(or alternatively spatial extent) by effective frequency width. In the
eighties, Daugman extended Gabor's work to two dimensional filters \cite{d80}.
For linear filters there is an ``uncertainty relation" which limits the
resolution simultaneously attainable in space and frequency. In the past
decade 2D Gabor functions have been applied to receptive fields of neurons
in the striate cortex by many authors. They concluded that this filters provide
a good description for the receptive field structure of simple cells in the
cat striate cortex. In the words of Daugman ``... the visual system is
concerned with extracting information jointly in the 2D space domain and
in the 2D frequency domain, and because of the incompatibility of these two
demands, has evolved towards the optimal solution via 2D channels that
roughly approximate 2D Gabor filters."  The problem of `energetic'
uncertainty principles
in human visual perception has recently been tackled by Trifonov and Ugolev
\cite{tu94}. Moreover, in their paper there is a good historical account of the
problem. The main                                                                                                                                                                                                                                              
                                                                                        a good historical account. The main
idea is that since the human eye responds to the emitted luminescence, one
may be endowed to look for an uncertainty principle involving the luminescence
threshold and the spatial resolution.

One can foresee that more complicated families of wavelets and wavelet-based
representations of the signals will be involved in reproducing the signal
processing of more complicated visual and auditory receptive fields of
neurons.
In this case, the detailed study of new types of uncertainty relations will be
of great importance. The interested reader is referred to the
literature \cite{b81}.

\subsection{Quantal synaptic transmission ?}

There is considerable debate in neurophysiology on the problem of quantal
synaptic transmission. This is a dominant hypothesis concerning the chemical
transmission, which is the principal means of neuronal communication in the
central nervous system. The debate centers around statistical analyses of
recorded histograms of excitatory postsynaptic currents, whose quantal nature
means demonstration of successive peaks, ideally evenly spaced, which are
thought to be
of biological and not of statistical origin \cite{lf92}. My opinion is that
whenever one is facing statistical treatment of data one should proceed with
extreme care since there may occur unexpected statistical artefacts. I agree
more with the demonstration provided by Clements \cite{cl91} that regularly
spaced peaks in a synaptic amplitude histogram can arise from sampling error
than with the answer of Larkman, Stratford, and Jack \cite{lsj91}.

\section{Limitations of the human brain to the quantum knowledge}

Recently, James D. Edmonds Jr. \cite{e11}
examined the human brain limitations
to quantum knowledge, citing Bohr's opinion that the task of physics is
to reveal what we can say about Nature and not what is Nature \cite{bh}.
 According to this
conjecture, which seems quite reasonable, ``we only do
brain-limited physics !''. Hence, our theories are only strategies, i.e.,
decision making in the face of uncertainties. However, the crucial
assumption which determines the structure of a strategy is due to
dynamics and not to probabilities and is based on microscopic
reversibility. This fundamental assumption gives rise to the equation
of detailed balance, which is, as a matter of fact, Bayes's postulate
in probability theory, i.e., the common way of conditioning for macroscopic
probabilistic thinking. It is well-known that microscopic reversibility does
not imply necessarily time reversal invariance \cite{pa93}. On the other hand,
the main components of logical reason are cause-effect relationships.
By their very nature cause-effect
correlations involve dynamics with a prefered direction of time.
It would be therefore interesting to develop non-Bayesian strategies,
since they might find a direct experimental field in the mesoscopic world.
Such strategies will
be applicable whenever one will take into account violations of
microscopic reversibility and the activity related to some mesoscopic
agent working like a Maxwell demon \cite{md1}. An interesting
discussion of the breakdown of microscopic reversibility in
enantiomorphous systems in the context of chemical evolution
and origins of life has been provided by L. D. Barron \cite{ldb},
who introduced the concept of enantiomeric detailed balancing, that can be
of interest to neuronal networks too.

The common logic of human thinking seems to be in difficulty whenever
probabilistic reasoning is coming into play.
It is not at all an easy matter to elaborate languages and appropriate
terminology for generalized probability judgements \cite{st85}. Indeed,
Arthur Miller attributed to Heisenberg the following remarkable recollection
of the years 1926-1927: ``we couldn't doubt that quantum mechanics was the
correct scheme but even then we didn't know how to talk about it, and the
discussions left us in a state of almost complete despair". As a matter of
fact we are at this point very close to the theories of language formation,
which predict a period of chaotic dynamics both in groups of cerebral neurons
and in the thalamocortical pacemaker \cite{nt89}. According to
Damasio \& Damasio \cite{dd92}: ``A large set of neural structures serves
to represent concepts; a smaller set forms words and sentences. Between the
two lies a crucial layer of mediation..." and I would say of ``meditation". It
is this layer of mediation that one can associate with the period of chaotic
dynamics.

At a more physical level, let's touch upon  Zipf's principle of minimal
effort in speaking \cite{z49} or, equivalently, Mandelbrot's condition of
minimal cost of information transmission \cite{m53}. Such variational
principles or conditions can be
associated with 1/f noise in speaking and writing as a manifestation of
information transmission in normal human communication. For a recent
derivation of a universal 1/f noise from an extremized physical information
see Frieden and Hughes \cite{fh94}. Recall now that a 1/f noise is only one
of the two requirements of the self organized criticality (SOC) paradigm. The 
second one is a fractal or
multifractal spatial structure of the region producing the 1/f noise, i.e.,
for speech, Broca's area, and for understanding languages, Wernicke's area.
What we suggest here is self organized critical states of these brain areas
as possible non-equilibrium dynamical brain states for normal verbal
communication. Passing to an electromagnetic (nonverbal) communication, and
accepting the idea of an electromagnetic lasing organ as alluded above, the
information
transmission would be through the vortex patterns in the transverse plane
of the laser beam \cite{bra91}, but again taking into account the result of
Frieden and Hughes \cite{fh94}, one can claim that a SOC paradigm will still
be at work, however at much superior levels of information rates.

An interesting debate concerns the non-verbality of thought ({\em Emperor's}
pp 423-425). There is the remarkable phrase of Henry Adams in his
{\em Education}: ``No one means all he says, and yet very few say all they
mean, for words are slippery and thought is viscous."
Many artists certainly don't think their masterpieces in words,
at least during the creative instants, and also a number of
eminent scientists were completely against words and insisted on their
drawback and even damaging effect with respect to thoughts (for examples, see
{\em Emperor's} pp 423-425). However, as Penrose mentions, there are persons
managing to process
a rapid and efficient transcription of their thoughts into words such as
philosophers, and this certainly with no less
merits. Admittedly, there are ways of thinking, like artistic and/or
scientific ones, for which words are not so much useful.
So what are thoughts really ?
Can they be associated with various transport phenomena of nerve signals, like
various types of solitons and other non-linear wave structures in neuronal
nets ? For example, one can work out a simple non-linear Schr\"odinger equation,
either discrete or continuous, for the propagation of thought interpreted
as an envelope soliton and discuss
``collapse"- like and/or ``blow-up"- like phenomena corresponding to various
phases of the
creativity processes. Moreover, non-linear extensions of the quantum mechanics,
not fulfilling the second law of thermodynamics \cite{a89},
may well be at their home inside the human brain, which being
a living system does not obey the usual formulation of the second law of
thermodynamics.

Penrose's discussion of the nerve signals
({\em Emperor's} pp 389-392) is very short. Hodgkin-Huxley
oscillator model and the FitzHugh-Nagumo one are two well-established nonlinear
models for this phenomenon. To fully be aware of the
importance of non-linear partial differential equations for pulselike
voltage waves carrying information along a nerve fiber I refer the reader to
the review paper of Scott \cite{s75}.

I also quote as being very close to Bohr's conjecture, Wolfram's
point of view
\cite{w11} who, in a cellular automaton context, claimed that physical
processes are only computations, whence the difficulty of answering
physical questions is directly connected to the difficulty of
performing the computations. At the quantum level of the human brain,
it will be of interest to obtain further insight into its ``quantum computer"
aspects \cite{d85},
taking into account the recent claims of improved efficiency for certain
algorithms \cite{u94}, and also for reasons implied by quantum logic theories
\cite{ql}.

\section{Conclusions}

In this essay, I expressed a range of speculative ideas that resulted from
the notes I used to make during my first reading of {\em The Emperor's New
Mind} and my simultaneous random jumping from shelf to shelf
in the ICTP-SISSA libraries. One warning for the reader is that none of those
ideas may be truly of worth, although my feeling is that human brain can
support phenomena described by generalized quantum methods, other than the
usual Ising-like transcription of memory patterns in neural networks.
Particularly interesting would be a generalized brain superradiance. 
Also direct vision (not by projections) in more than three dimensions is another 
interesting issue.

Quantum mechanics {\em per se} seems to be a weak theory and not a proper
scientific language when confronting it
with the complexity of the brain activity, and also when compared with other
methods put forth in tackling this highly interdisciplinary research field.
However, the progress in our
technologies and the advancement of our understanding of the functioning of
the human brain at quantum and mesoscopic levels may well have important
consequences in the future. It is somewhat amasing yet not surprising,
that while the most
advanced tomographic techniques of visualising the brain activity are based
on quantum mechanical phenomena, we have so little to say about the
quantum-mechanical brain.
For the time being, the main doctrine that
brain activity is entirely computation is dominating the field despite a few
metaphysical objections related for example to the consciousness issue,
and I am afraid that even the microtubules and their infrastructure can be
included in a computational scheme (according to the principle that digital
computing can be used to model and/or to describe most physical systems).
Indeed, M.P. Barnett \cite{b87} has already suggested that
microtubules are processing channels along which strings of bits are
propagating from one place to another, and they may well be the material
base for the {\em ultimate computing} \cite{h87} in the molecular framework.
Microtubule networks may turn into a major
research field in the near future. For example, they are predicted to
possess piezoelectric properties allowing a possible application of
recently proposed experimental techniques called two-photon diffraction
and holography \cite{bk94}.

Finally, whether or not the quantum features of the human brain will prove
difficult to reveal, this does not mean at all that a quantum brain cannot be
fabricated.





\begin{center}
{\bf More references}
\end{center}

\vskip 0.3cm

\begin{itemize}
\item {\bf Consciousness, Microtubules, Robot, Thought, Quantum Mechanics,
Perception}
\end{itemize}

I. Marshall,
``Consciousness and Bose-Einstein condensates", New Ideas in
Psychol.{\bf 7}, 73-83 (1989)

P. P\'erez,
``Physics and Consciousness", quant-ph/9510017

F. Crick and C. Koch,
``Towards a neurobiological theory of consciousness",
{\em seminars in The Neurosciences} {\bf 2}, 263-275 (1990)
[visual awareness (VA)- the favorable form of consciousness to study
neurobiologically is proposed to be made of a very fast form associated to
iconic memory whose neural basis is difficult to study experimentally,
and a slower form
based on visual attention and short-term memory. In the slower VA, the iconic
memory is due to an attention mechanism which binds togheter a cluster of
neurons whose activity is related to the main features of the visual
object on which the attention is paid. It is suggested that the
cluster's collective activity is done by generating coherent semi-synchronous
oscillations, probably in the 40-70 Hz range. These oscillations are the
source of a transient short-term memory. Several lines of experimental work
are outlined that might help to understand the neural mechanism involved.]

Lester Ingber,
``Statistical mechanics of neocortical interactions: Constraints on 40-Hz
models of short-term memory", Phys. Rev. E {\bf 52}, 4561-63 (1995)

J.E. LeDoux,
``Emotion, memory and the brain",
Scientific American , June issue of 1994

D.N. Page,
``Attaching theories of consciousness to Bohmian quantum
mechanics", quant-ph/9507006
[theories of conscious perceptions are attached to Bohmian QM turning it into
Sensible Bohmian QM. Previously, Don Page introduced the SQM concept, see
``SQM: Are only perceptions probabilistic", quant-ph/9506010 and references
therein.]

G. Vitiello,
``Dissipation and brain", and the extended version with the title
``Dissipation and memory capacity in the quantum brain model",
Int. J. Mod. Phys. B {\bf 9} (1995)
[the quantum model of the brain proposed by Ricciardi and Umezawa is
extended to dissipative dynamics. Consciousness emerges as a manifestation
of the brain dissipative dynamics.]

L.M. Ricciardi and H. Umezawa,
Kibernetic {\bf 4}, 44 (1967)

V.E. Tarasov,
``Strings and dissipative mechanics", hep-th/9506053

D.V. Nanopoulos,
``Theory of brain function, quantum mechanics and
superstring", hep-ph/9505374
[Fig. 1: Psychological or
Personality profile as a function of time, parametrized in terms of
the so-called synchordic collapse frequency $\gamma \equiv 1/\tau _c$
of the microtubule-network.]

N.E. Mavromatos and D.V. Nanopoulos,
``Non-critical string theory formulation of MT dynamics and quantum aspects
of brain function", ACT-09/95, CTP-TAMU-24/95, ENSLAPP-A-524/95;
``On a possible connection of non-critical strings to certain aspects of
quantum brain function", quant-ph/9510003


R. Penrose and S. Hameroff,
``Orchestrated reduction of quantum coherence in brain MTs:
A model of Consciousness", Mathematics and Computers in Simulation
{\bf 40}, 453-480 (1996)

J.W. Griffin et al.,in
{\em Molecular biology of the
human brain} (1988 Alan R. Liss, Inc.), pp 141-145.
[Excerpt: MTs in neurons, as in other cells, are composed of alpha and beta
subunits.
MTs are found throughout the neuron. They are the major cytoskeletonal
element in dendrites. In the axon, they are arranged as parallel, strictly
longitudinal rods, with the length of individual MTs estimated to be
350-750 microns in length. They are uniformly polarized with the minus
end toward the periphery....
MTs provide the essential structural substrate for fast axonal transport
within the axon.]

H.P. Stapp,
``Why classical mechanics cannot naturally accommodate
consciousness but quantum mechanics can", LBL-36574 (February 8, 1995),
prepared for a special issue of {\em Psyche}.
``Quantum mechanical coherence, resonance, and mind", quant-ph/9504003;
``Values and the quantum conception of man", quant-ph/9506035;
``The hard problem: A quantum approach", LBL-37163 (June 1, 1995).
``Quantum propensities and the brain-mind connection", Found. Phys. {\bf 21},
1451 (1991)

J.C. Eccles,
``Do mental events cause neural events analogously to the
probability fields of quantum mechanics ?", Proc. R. Soc. Biol. {\bf 227},
411-428 (1986)

D.C. Bennett,
``The practical requirements for making a conscious robot",
Phil. Trans. R. Soc. Lond. A {\bf 349}, 133 (1994)
[longterm project to design and build a humanoid {\em protein-free}
conscious robot named Cog]

W.A.N. Ellis,
``A theory of thought", Spec. in science and techn. {\bf 16}, 181 (1993)

A. Rose,
``Quantum effects in human vision", in Adv. in Biol. and Medical
Physics {\bf 5}, 211-242 (1957)

J.A. Wheeler and W.H. Zurek (eds),
{\em Quantum theory and measurement}, Princeton series in Physics,
(Princeton University Press, Princeton, New Jersey, 1983), section
{\em Questions of principle} at pp. 769-771.

D.L. Gonzalez et al.,
``Pitch perception of complex sounds: Nonlinearity revisited", chao-dyn/9505001

\vskip 0.5cm
\begin{itemize}
\item {\bf The Neuron: Some experiments}
\end{itemize}

P. Fromherz and A. Stett,
``Silicon-Neuron junction: Capacitive stimulation
of an individual neuron on a silicon chip", PRL {\bf 75}, 1670 (1995)

G. De Stasio et al.,
``Scanning photoemission spectromicroscopy of neurons",
PR E {\bf 48}, 1478 (1993)

R. Uma Maheswari et al.,
``Observation of subcellular nanostructure of
single neurons with an illumination mode photon scanning tunneling
microscope", Opt. Commun. {\bf 120}, 325-334 (1995)

\vskip 0.5cm

\begin{itemize}
\item {\bf Stochastic Resonance, Complexity, Firing, Noise, Chaos,
Fractal, Dendritic Tree, Neural Computation}
\end{itemize}

P.C. Bressloff,
``Average firing rate of a neural network with dynamical
 disorder", J. Phys. A {\bf 28}, 2457-2469 (1995)

P. Jung,
``Threshold devices: Fractal noise and neural talk",
Phys. Rev. E {\bf 50}, 2513 (1994)

L.A. Lipsitz and A.L. Goldberger,
``Loss of `complexity' and aging",
J. Am. Med. Association  (JAMA) {\bf 267}, 1806 (1992).
[Fig. 3 in this paper, which is reprinted with permission from WB Saunders Co.,
shows an age-related loss of fractal structure in the dendritic arbor of
giant pyramidal Betz cell of the motor cortex.]

T. Fukai,
``Metastable states of neural networks incorporating the physiological Dale
hypothesis", J. Phys. A {\bf 23}, 249 (1990)

J.A. Scott Kelso and A. Fuchs,
``Self-organizing dynamics of the human brain: Critical instabilities and
Sil'nikov chaos", Chaos {\bf 5}, 64-69 (1995) and references therein.

S.J. Schiff et al.,
``Controlling chaos in the brain", Nature {\bf 370}, 615 (1994)

J.K. Douglass, L. Wilkens, E. Pantazelou and F. Moss,
``Noise enhancement of information transfer in
crayfish mechanoreceptors by SR", Nature {\bf 365}, 337 (1993)
[SR observed in experiments with hydrodynamically sensitive hair
mechanoreceptors in the tailfan of the crayfish
{\em Procambarus clarkii}- termed also
crayfish neuron experiment]

A. Longtin, A. Bulsara and F. Moss,
``Time-interval sequences in bistable
systems and the noise-induced transmission of information by sensory
neurons", Phys. Rev. Lett. {\bf 67}, 656 (1991)
[simulations of the effect of the noise intensity on the interspike interval
histograms]

A.R. Bulsara, R.D. Boss, and E.W. Jacobs,
``Noise effects in an electronic
model of a single neuron", Biol. Cybern. {\bf 61}, 211 (1989)

A.R. Bulsara, A.J. Maren, G. Schmera,
``Single effective neuron: dendritic
coupling effects and SR", Biol. Cybern. {\bf 70}, 145-56 (1993)

A. Longtin, A. Bulsara and F. Moss,
``Sensory information processing by noisy bistable neurons",
Mod. Phys. Lett. B {\bf 6}, 1299 (1992)

A.R. Bulsara, S.B. Lowen and C.D. Rees,
``Cooperative behavior in the periodically modulated Wiener process:
Noise-induced complexity in a model neuron", Phys. Rev. E {\bf 49}, 4989 (1994)

K. Wiesenfeld,
``SR on a circle", Phys. Rev. Lett. {\bf 72}, 2125 (1994)
[electronic Fitzhugh-Nagumo neuron model, which essentially describes a
threshold-plus-reinjection dynamics].

M.C. Teich,
``Fractal neuronal firing patterns", in {\em Single neuron
computation. Neural nets: Foundations to applications} (Academic Press, 1992)
pp. 589-625.

J.G. Elias,
``Artificial dendritic trees", Neural Computation {\bf 5}, 648
(1993)

P. Ling,
``Neurocomputation by reaction diffusion", PRL {\bf 75}, 1863 (1995)

\'A. T\'oth and K. Showalter,
``Logic gates in excitable media",
J. Chem. Phys. {\bf 103}, 2058 (1995)

N. Suga,
``Biosonar and neural computations in bats", Scientific Am. (June 1990)

J.A. Simmons, D.J. Howdell and N. Suga,
``Information content of bat sonar echoes",
Am. Scientist {\bf 63}, 204 (1975)

H.C. Tuckwell,
{\em Stochastic processes in the neurosciences}, 
(SIAM, Philadelphia, 1989)

\vskip 0.5cm

\begin{itemize}
\item {\bf Laser-type phenomena}
\end{itemize}

G. Alli and G.L. Sewell,
``New methods and structures in the theory of the
multimode Dicke laser model",  J. Math. Phys. {\bf 36}, 5598 (1995)

S.T. Zavtrak,
``Generation conditions for an acoustic laser",
Phys. Rev. E {\bf 51}, 3767-69 (1995)

\vskip 0.5cm

\begin{itemize}
\item {\bf Encephalograms}
\end{itemize}

N. Pradhan and P.K. Sadasivan,
``Relevance of surrogate-data testing in electroencephalogram analysis",
Phys. Rev. E {\bf 53}, 2684 (1996)



\end{document}